\documentstyle[twoside,fleqn,espcrc2,epsf]{article}

\newcommand{\Z}{{\sf Z \!\!\! Z}}

\newcommand{\AmS}{{\protect\the\textfont2
  A\kern-.1667em\lower.5ex\hbox{M}\kern-.125emS}}

\title{QCD strings ending on domain walls --- a complete wetting phenomenon
in SUSY QCD}

\author{A.Campos, K. Holland 
	and
	U.-J. Wiese\address{Center for Theoretical Physics,
        Laboratory for Nuclear Science and Department of Physics, \\
        Massachusetts Institute of Technology (MIT), 
	Cambridge, Massachusetts 02139, U.S.A.}
	\thanks{Talk presented by K.H. This work is supported in 
        part by funds provided by the U.S. Department of Energy 
        (D.O.E.) under cooperative agreement DE-FC02-94ER40818}} 

\begin{document}

\begin{abstract}
In the context of M-theory, Witten has argued that an intriguing phenomenon
occurs, namely that QCD strings can end on domain walls. We present a 
simpler explanation of this effect using effective field theory to 
describe the behavior of the Polyakov loop and the gluino condensate in
${\cal{N}}=1$ supersymmetric QCD. We describe how domain walls separating 
distinct confined phases appear in this effective theory and how these 
interfaces are completely wet by a film of deconfined phase at the 
high-temperature phase transition. This gives the Polyakov loop a 
non-zero expectation value on the domain wall. Consequently, a static test 
quark which is close to the interface has a finite free energy and the 
string emanating from it can end on the wall.   

\end{abstract}

\maketitle

\section{SYMMETRIES}

Witten's argument that a QCD string can end on a domain wall is made in
the context of M-theory \cite{Wit97}. He shows that this can occur in a 
low energy limit of a theory in the universality class of ${\cal{N}}=1$ 
supersymmetric (SUSY) $SU(N)$ Yang-Mills theory. This is a theory which 
contains gluons and their supersymmetric partners, gluinos, all of which 
transform under the adjoint representation of the gauge group. However, 
the concepts of QCD strings and domain walls naturally occur in  
field theory and it seems natural to explore if a field 
theoretical explanation of this effect could be found. There has been much 
investigation into domain walls in SUSY Yang-Mills theories at zero 
temperature \cite{Kov97}. We begin with ${\cal{N}}=1$ SUSY $SU(N)$ 
Yang-Mills theory, extract the features which we think are relevant to this 
phenomenon and construct an effective theory at non-zero temperature. 
We will want to know if our effective theory
contains QCD strings, which is essentially equivalent to whether or not
fundamental charges are confined. From the microscopic theory, we can 
construct the Polyakov loop $\Phi(\vec{x})$ whose expectation value gives us
the free energy $F$ of a static test quark placed in our system, i.e.
$\langle \Phi \rangle \propto \exp(-F/T)$, where $T$ is the temperature of the
system. If the quark is confined and $F$ diverges, then $\langle \Phi 
\rangle$ vanishes. Alternatively, if 
$\langle \Phi \rangle$ is non-zero, then $F$ is finite and the quark is 
deconfined. The
microscopic theory has at non-zero temperature a discrete symmetry called
the ${\Z}(N)_c$ center symmetry, which is due to gauge transformations
being not quite periodic in the Euclidean time direction. A non-zero
value for the Polyakov loop, the signal for deconfinement, spontaneously 
breaks this center symmetry. 
For domain walls to occur, a global discrete symmetry must be spontaneously
broken. The microscopic SUSY Yang-Mills theory also
contains a discrete ${\Z}(N)_{\chi}$ chiral symmetry, which is a 
remnant of
the anomalously broken $U(1)$ axial symmetry. This discrete chiral symmetry
of the full quantum theory is broken by a non-zero expectation value for
the gluino condensate $\chi$. If this symmetry is spontaneously broken,
domain walls appear which separate regions where the system has fallen into
one of the $N$ distinct vacua, distinguished by different phases for the
complex-valued gluino condensate. Our effective field theory contains the 
Polyakov loop and the gluino condensate and satisfies the ${\Z}(N)_c$
center and ${\Z}(N)_{\chi}$ chiral symmetries and is also invariant under
charge conjugation.

\section{EFFECTIVE THEORY}

We construct an effective theory for supersymmetric Yang-Mills theory
with the gauge group $SU(3)$ with the effective action 
\begin{eqnarray}
S[\Phi,\chi] = \int d^3 x \left[ \frac{1}{2}|\partial_i \Phi|^2 +
\frac{1}{2}|\partial_i \chi|^2 + V(\Phi,\chi) \right]. \nonumber
\end{eqnarray}
We demand that the potential $V$ be invariant under ${\Z}(3)_c$ center,
${\Z}(3)_{\chi}$ chiral and charge conjugation transformations. Under
${\Z}(3)_c$ transformations, $\Phi \rightarrow \Phi z$ and $\chi \rightarrow 
\chi$, where $z \in {\Z}(3)_c = \{ e^{2 \pi i n/3}, n = 0,1,2 \}$. Under
${\Z}(3)_{\chi}$ transformations, $\chi \rightarrow \chi z^{'}$ and $\Phi
\rightarrow \Phi$, where $z^{'} \in {\Z}(3)_{\chi} = \{ e^{2 \pi i m/3}, 
m = 0,1,2 \}$. Under charge conjugation, $\Phi \rightarrow \Phi^{*}$ and  
$\chi \rightarrow \chi^{*}$. We are
interested in the universal properties of all potentials $V$ which obey
these symmetries. We do not need to calculate the exact potential which
is obtained by integrating out all degrees of freedom other than $\Phi$
and $\chi$. The simplest potential we can write down that has these 
properties is
\begin{eqnarray}
&& \hspace{-.3in} V(\Phi,\chi) = a |\Phi|^2 
+ b \Phi_{1}(\Phi_{1}^{2} - 3 \Phi_{2}^{2})
+ c |\Phi|^4 \nonumber \\ 
&& \hspace{-.1in} + \ d|\chi|^2 + e \chi_{1}(\chi_{1}^{2} - 3 \chi_{2}^{2}) 
+ f |\chi|^4  + g |\Phi|^2 |\chi|^2, \nonumber
\end{eqnarray}
where $\Phi = \Phi_{1} + i \Phi_{2}$ and $\chi = \chi_{1} + i \chi_{2}$.
At low temperatures, we know that the gluons and gluinos are confined
and that the discrete chiral symmetry is broken. At high temperatures,
the center symmetry is spontaneously broken, signalling deconfinement, and 
the chiral symmetry is restored. We assume that there is one first order
phase transition, where the ${\Z}(3)_c$ symmetry is spontaneously broken 
and simultaneously the ${\Z}(3)_{\chi}$ symmetry is restored.
This means that we have six types of bulk phase --- three kinds of confined
phase with $\Phi^{(n)} = 0$ and $\chi^{(n)} = \chi_{0} e^{2 \pi i n/3}, 
n = 0,1,2$ and three kinds of deconfined phase with $\chi^{(n)} = 0$ and
$\Phi^{(n)} = \Phi_{0} e^{2 \pi i n/3}, n = 3,4,5$. The temperature of the
system is contained implicitly in the parameters $a,b,...,g$ in the potential
$V$. The phase transition temperature corresponds to a choice of parameters
where the six bulk phases are degenerate absolute minima of the potential.
We look for planar domain wall solutions of the classical equations of motion
so $\Phi(\vec{x})=\Phi(z)$ and $\chi(\vec{x})=\chi(z)$, where $z$ is the 
coordinate perpendicular to the wall. The equations of motion are
\begin{eqnarray}
\frac{d^{2} \Phi_{i}}{d z^{2}} = \frac{\partial V}{\partial \Phi_{i}} , 
\frac{d^{2} \chi_{i}}{d z^{2}} = \frac{\partial V}{\partial \chi_{i}}.
\nonumber
\end{eqnarray}
\begin{figure}[tb]
\vspace*{-5mm}
\centerline{
\epsfxsize=8cm\epsfbox{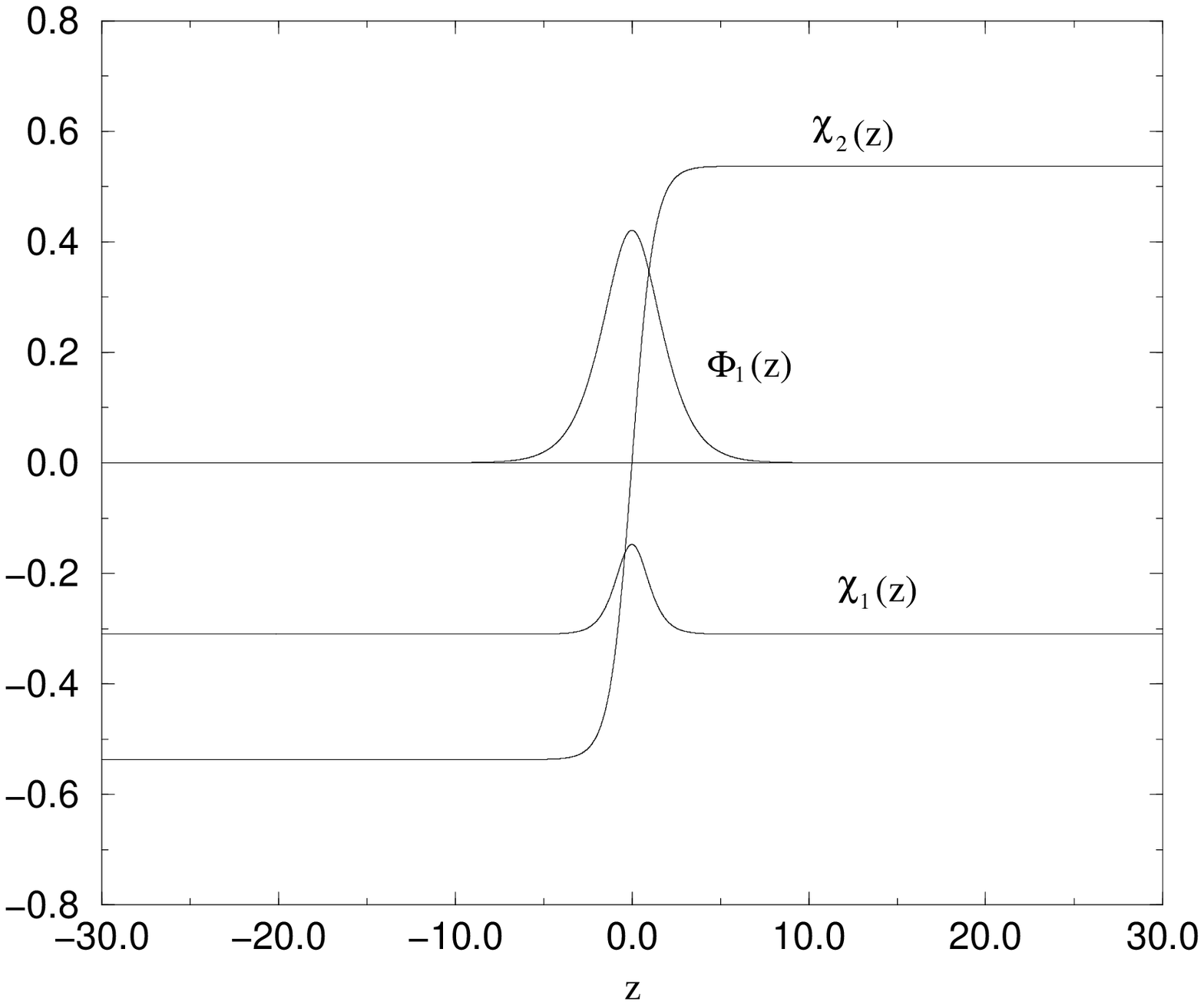}
}
\vspace*{-5mm}
\centerline{
\epsfxsize=8cm\epsfbox{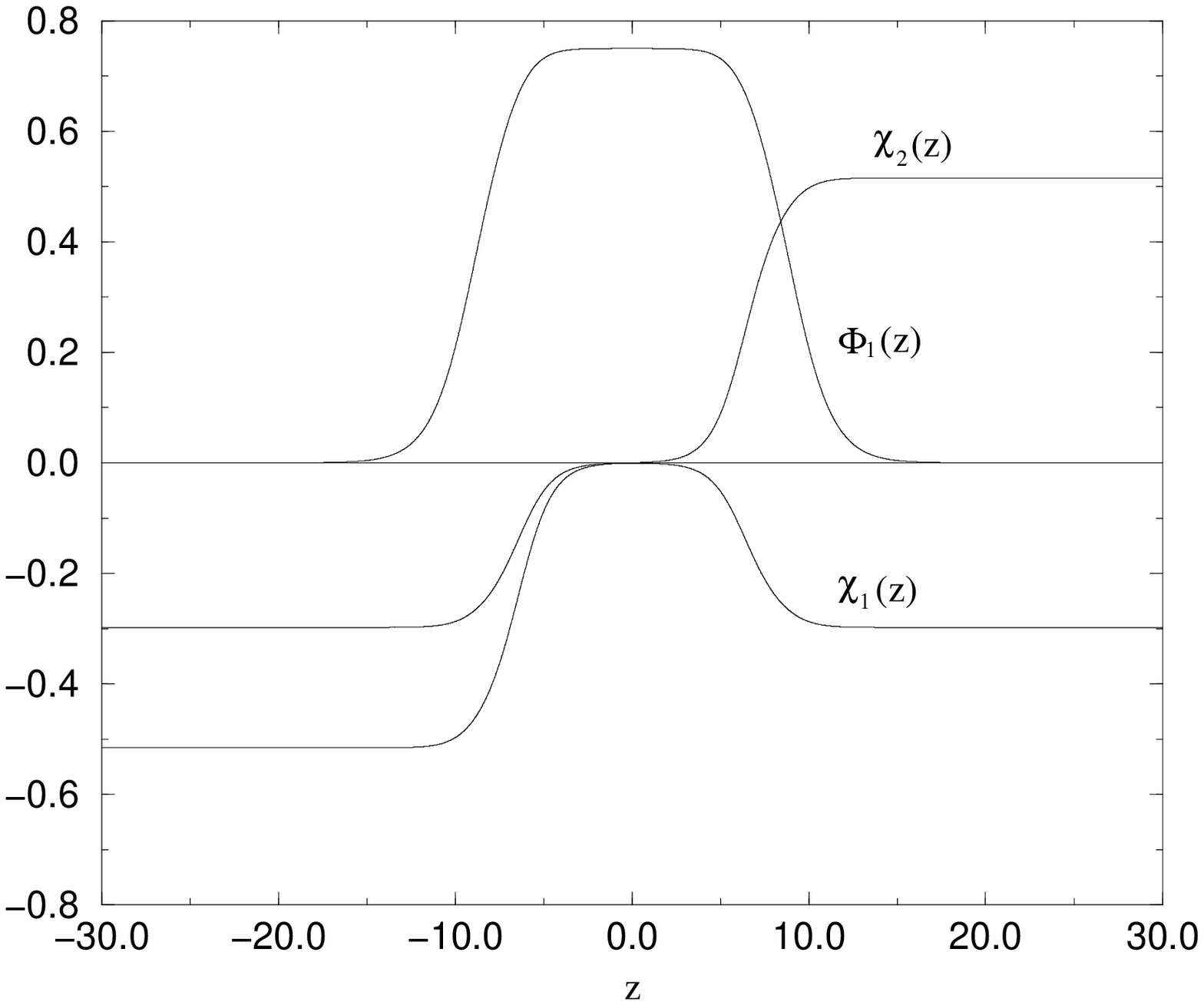}
}
\vspace*{-10mm}
\caption{
Profile of confined-confined domain wall (a) deep in the confined regime far
from the phase transition and (b) close to the critical temperature.
}
\label{fig:wetting}
\end{figure}
We look for a numerical solution representing a domain wall separating 
two regions of confined bulk phase, one of type $(1)$ and the other of 
type $(2)$, i.e. we apply
the boundary conditions $\Phi(\infty) = \Phi^{(1)}, \chi(\infty) = \chi^{(1)}$
and $\Phi(-\infty) = \Phi^{(2)}, \chi(-\infty) = \chi^{(2)}$. In Figure 1(a),
we show a solution deep in the confined regime, far from the critical
temperature. However, we see that the Polyakov loop has a non-zero value
on the domain wall, showing that the interface has some of the properties of
the deconfined bulk phase. In Figure 1(b), we show a solution at a temperature
close to the phase transition. The confined-confined interface has now split
into two confined-deconfined interfaces, which are separated by a complete
wetting layer of deconfined bulk phase. Complete wetting is a universal
phenomenon described by a set of critical exponents determined by the range
of the interaction between the two interfaces. For example, we find that
the width of the wetting layer grows logarithmically as $z_{0} \propto
- \ln (T-T_{c})$, a behavior predicted for short-range forces \cite{Lip84}. 
We have found critical behavior in other quantities, e.g. the interface 
tension of the two confined-deconfined interfaces, and have also found an
analytic solution to the equations of motion \cite{Cam98}. We have
also investigated in detail the occurence of complete wetting at zero
temperature in systems with confined and Coulomb bulk phases \cite{Hol98}.
Complete wetting
is a very general phenomenon and was conjectured and found to occur also in 
non-supersymmetric $SU(3)$ pure gauge theory \cite{Fre89}.

However, the most important point to be extracted from the solutions is that
even at a temperature deep in the confined regime, the domain wall has a
non-zero Polyakov loop expectation value. This means that a static quark 
placed at the domain wall has a finite free energy --- the infinite thin film 
of deconfined phase at the interface acts as a sink for the color flux 
emanating from the quark. Using the potential $V$, we can also solve for
$\Phi$ near the interface and we find that the quark's free energy $F$ 
grows linearly with the
distance between the quark and the domain wall. This is exactly what we would
expect for a QCD string connecting the quark to the domain wall, where the
energy cost is proportional to the string length. The 2-d complete wetting film
can transport the color flux to infinity at a finite energy cost because
of Debye screening. We assumed that deconfinement and chiral symmetry 
restoration occur at the same temperature. Starting at low temperature, if 
instead chiral symmetry is restored before deconfinement occurs, then we 
would again have complete wetting. However, the wetting layer would be chirally
symmetric confined phase and so a quark placed at the domain wall would
still have an infinite free energy --- a QCD string could not end there.
Alternatively, if deconfinement happens before chirally symmetry restoration,
we have a new phenomenon called incomplete wetting. Instead of the infinite 
wetting layer, bubbles of deconfined phase form at the confined-confined 
interface. Because these bubbles have finite volume, quantum tunneling
between the various kinds of deconfined phase bubbles means that the 
functional integral over all domain wall solutions gives $\langle \Phi
\rangle = 0$, so again a quark placed at the interface has an infinite free
energy. This is consistent with Gauss' law which says that the string cannot
end on a bubble as it has a compact surface. Quantum tunneling does not occur
between infinite films of distinct types of deconfined phase as these 
solutions are separated by infinite energy barriers. Note that it is essential
that the domain wall be infinite so that the color flux can be transported 
to infinity. If we believe Witten's argument that QCD strings can end on 
domain walls, then we can make a prediction for the phase structure of 
$SU(N)$ Yang-Mills theory with gluons and gluinos. We conclude that in this
model chiral symmetry is restored and deconfinement occurs at the same 
temperature.

\end{document}